\documentclass[prb,preprint,showpacs,superscriptaddress]{revtex4}
\usepackage{graphicx}
\usepackage{amsmath}
\usepackage{amssymb}
\usepackage{color}
\begin{document}
\preprint{PREPRINT (\today)}

\title{Clean and Dirty Superconductivity in Pure, Al doped, and Neutron Irradiated
MgB$_{2}$: a Far-Infrared Study}

\author{M. Ortolani}
\affiliation{Coherentia-INFM and Physics Department, University of
Rome ``La Sapienza'', P.le A. Moro 5, I-00185, Rome, Italy}

\author{D. Di Castro}
\affiliation{Coherentia-INFM and Physics Department, University of
Rome ``La Sapienza'', P.le A. Moro 5, I-00185, Rome, Italy}
\affiliation{Physics Institute, University of Zurich, Winterthurerstr. 190, CH-8057, Zurich, Switzerland}

\author{P. Postorino}
\affiliation{Coherentia-INFM and Physics Department, University of
Rome ``La Sapienza'', P.le A. Moro 5, I-00185, Rome, Italy}

\author{I. Pallecchi}
\affiliation{INFM-LAMIA, Dipartimento di Fisica, Via Dodecaneso
33, 16146 Genova, Italy}

\author{M. Monni}
\affiliation{INFM-LAMIA, Dipartimento di Fisica, Via Dodecaneso
33, 16146 Genova, Italy}

\author{M. Putti}
\affiliation{INFM-LAMIA, Dipartimento di Fisica, Via Dodecaneso
33, 16146 Genova, Italy}

\author{P. Dore}
\affiliation{Coherentia-INFM and Physics Department, University of
Rome ``La Sapienza'', P.le A. Moro 5, I-00185, Rome, Italy}

\begin{abstract}
The effects of Al substitution and neutron irradiation on the
conduction regime (clean or dirty) of the $\pi$- and $\sigma$-band of MgB$_{2}$
have been investigated by means of far-infrared spectroscopy. The
intensity reflected by well characterized polycrystalline samples
was measured  up to 100 cm$^{ - 1}$ in both  normal and
superconducting state. The analysis of the superconducting to
normal reflectivity ratios shows that only the effect of the opening of the small gap in the dirty
$\pi$-band can be clearly observed in pure MgB$_{2}$, consistently with previous
results. In Al-doped samples the dirty character of the $\pi$-band
is increased, while no definitive conclusion on the conduction
regime of the $\sigma $-band can be drawn. On the contrary,
results obtained for the irradiated sample show that the
irradiation-induced disorder drives the $\sigma$-band in the dirty
regime, making the large gap in $\sigma$-band observable for the first
time in far-infrared measurements.
\end{abstract}

\maketitle

A large number of theoretical \cite{liu,choi} and
experimental \cite{bouq, gonn, quil} works showed that MgB$_{2}$
is a two-band superconductor with two distinct energy gaps: a
small gap $\Delta _{\pi}$ originating from the three-dimensional
(3D) $\pi $-bands, formed by $p_{z}$ orbitals of boron along the
$c$-axis, and a large gap $\Delta _{\sigma }$ originating from the
two-dimensional (2D) $\sigma $-bands in the \textit{ab}-plane,
formed by the boron \textit{sp}$^{2}$ hybrids orbitals. Different
measurements showed that $\Delta _{\pi}$ and $\Delta _{\sigma}$
are in the range 1.5-3.5 meV and 6.0-7.5 meV,
respectively.\cite{bouq, gonn, quil} Substitutional impurities
introduced in the system can modify the \textit{intraband
}impurity scattering in the $\pi $ and/or $\sigma $ bands, thus
affecting their transport properties. On the contrary, an increase of the
small \textit{interband }scattering between $\pi $ and $\sigma $
bands is difficult to achieve, owing to the different parity of
$\pi $ and $\sigma $ orbitals. \cite{mazin} A large number of
studies has then been devoted to investigate the effects of
substitutional impurities in MgB$_{2}$, in particular Al for Mg
\cite{slus,bianc} and C for B, \cite{mas,wilke,karp} which allows
to selectively disorder $\pi $ or $\sigma $ bands, thus giving the
unique opportunity to tune normal and superconducting state
properties.

In the case of a conventional (i.e., single-band s-wave) BCS
superconductor, characterized by coherence length $\xi_{0}$ and
energy gap $\Delta$, the decrease of the electron mean free-path
$\ell$ due to the increase of impurity scattering modifies the
conduction regime, which is classified in the clean-limit when $\ell >
\xi_{0}$ and in the dirty-limit when $\ell < \xi_{0}$. In studying
the conduction regime, infrared spectroscopy can be a powerful
tool for investigating important physical quantities such as the
gap $\Delta$, the plasma frequency $\Omega$, and the scattering
rate $\Gamma$. The optical response of the system is indeed given
by the complex optical conductivity
$\sigma(\omega)=\sigma_1(\omega)+i\sigma_2(\omega)$. In
the normal state, $\sigma_{N}$ is determined by $\Omega $ and
$\Gamma $, accordingly to the standard Drude model. In the
superconducting state, accordingly to the Mattis and Bardeen (MB)
model, the ratio $\sigma_{S}/\sigma_{N}$ is determined by
$\Delta$, since, for 0$<\omega<2\Delta / \hbar$,
$\sigma_{1,S}(\omega) = 0$ and $\sigma_{2,S}(\omega) \propto
1/\omega$. \cite{tink0} Therefore the reflectance spectrum $R(\omega)$ of a
bulk superconductor monotonically increases towards 1 as
$\omega$ goes to 0 in the normal state ($R_{N})$, while it steeply
increases to 1 as $\omega$ goes to $2\Delta / \hbar $ in the
superconducting state ($R_{S})$. Therefore, an increase in the $R_{S}/R_{N}$ reflectivity ratio, followed by a maximum around $\omega = 2\Delta / \hbar $, is
expected on decreasing frequency. \cite{tink0, timus}
The same arguments explain the
maximum observed in the superconducting to normal transmission
ratio obtained on thin superconducting films. \cite{tink1, will}
 However, this behavior strongly depends on the conduction regime. In particular, when
$\Gamma \ll 2\Delta / \hbar$, the maximum disappears since, in the
normal state, $\sigma_{1,S}(\omega)$ is already close to zero at
$\omega =2\Delta / \hbar $. \cite{kam} It is worth to notice that,
in optics, the condition $\Gamma \ll 2\Delta / \hbar$ ($\Gamma \gg
2\Delta / \hbar$), which correspond to $\ell \gg \xi_0$ ($\ell \ll
\xi_0$) since $\ell=v_{F}/\Gamma $ and $\xi _0 = \hbar v_F / \pi
\Delta $, is conveniently introduced in defining the clean (dirty)
limit condition. \cite{timus} In the two-band superconductor
MgB$_{2}$, the situation is far more complex because of the
different characteristics of $\pi$- and $\sigma$-bands. In
particular, recent Raman measurements \cite{quil} indicate that
$\pi $ and $\sigma $ carriers are in the dirty and clean regime,
respectively. This result is supported by trasmittance
measurements on MgB$_{2}$ thin films \cite{kaind, jung} and
reflectivity measurements on MgB$_{2}$ films \cite{pimen} and
single crystals, \cite{peruc} where the effect of  the small
$\Delta_{\pi}$ gap (dirty limit) is observed, while no effect of
the $\Delta_{\sigma}$ gap (clean limit) is revealed.

The aim of the present work was to monitor, by means of far-infrared (FIR)
spectroscopy, the effects of disorder induced by Al substitution
or neutron irradiation on $\Gamma$ and $\Delta$, i.e. on the
conduction regime, of the MgB$_{2}$ $\pi$- and $\sigma$-band.
Reflectivity measurements were thus performed in the FIR region on
pure MgB$_{2}$ (sample A), Mg$_{0.85}$Al$_{0.15}$B$_{2}$ (B), and
Mg$_{0.7}$Al$_{0.3}$B$_{2}$ (C) polycrystalline samples. Details
on sample preparation and characterization are reported elsewhere.
\cite{put1} A fourth sample (D), obtained by irradiating
Mg$^{11}$B$_{2}$ with a thermal neutron fluence of 10$^{18}$
n/cm$^{2}\,$, \cite{put2} was also measured.
The values of the parameters which characterize the investigated samples are
reported in Table I. $T_{C}$ is strongly decreased by Al doping,
in agreement with previous reports, \cite{slus, bianc} while it is
only slightly reduced by irradiation. The resistivity $\rho $ at
40 K (\textit{$\rho $}$_{40K})$ is as low as 2.2 $\mu \Omega $cm
for the pure sample, and it increases up to 24 $\mu \Omega $cm
with Al doping, and to 20 $\mu \Omega $cm with irradiation. The
energy gaps $\Delta _{\sigma}$ and $\Delta _{\pi}$ were measured
only for the A, B, and C samples by means of point-contact
spectroscopy. \cite{put1} By using \textit{ab-initio} electronic
band-structure computations, \cite{prof,put1} the doping-dependent
anisotropic plasma frequencies ($\Omega_{\sigma} ^{(a,b)},
\Omega_{\sigma}^{(c)}, \Omega_{\pi} ^{(a,b)}$, and
$\Omega_{\pi}^{(c)}$) were obtained. It is worth to notice that, for
pure MgB$_{2}$, the quoted $\Omega$ values are well consistent
with previous determinations. \cite{liu}

Reflectivity measurements were performed in the FIR region up to
about 100 cm$^{-1}$ with a rapid-scanning Bomem DA3 Michelson
interferometer, equipped with a liquid-cooled stabilized Hg lamp,
a bolometer cooled at 1.6 K, and different Mylar beamsplitters.
Each sample was mounted on the cold head of a cryostat equipped
with a temperature controller which allows long-time data
acquisition down to 8 K at temperatures stabilized within 0.5 K.
The large surface of the samples (around 6 mm in diameter) had an
appreciable reflectivity in the visible region although in
polycrystalline form. This allowed to carefully align the
sample, and to collect the reflected intensity at near-normal
incidence using a large dimension IR beam. For each sample, measurements were carried out  at 8 K
($R_{S})$ and at 45 K ($R_{N})$, in order to directly obtain the
\textit{RR= R}$_{S}$(8 K)/$R_{N}$(45 K) reflectivity ratio. This
procedure, \cite{gorsh} which does not need reference sample or
gold evaporation at room temperature to measure the absolute
reflectance, allows to avoid systematic errors due to
misalignments of the sample or to additional thermal cycles. Each
 spectrum was obtained by collecting 5000 interferograms, within an
acquisition time of at least 1 hour. The obtained \textit{RR}
ratios are shown in Figure 1 as a function of the wavenumber
$\nu$. By considering reproducibility of repeated measurements,
the uncertainties on the \textit{RR} ratios, mainly due to lamp
instability and Fourier transformations, were estimated to be $\pm
$0.005 and $\pm $0.002 at 30 cm$^{ - 1}$ and 70 cm$^{ - 1}$,
respectively. Low-frequency \textit{RR} data on which the
uncertainty was larger than 0.005 were discarded.

A first inspection of Fig. 1 shows that for the pure MgB$_{2}$
sample (panel a) the \textit{RR} ratio clearly increases below the
frequency-edge around 50 cm$^{ - 1}$ ($ \simeq $6 meV), in
agreement with a previous FIR study on polycrystalline
MgB$_{2}$.\cite{gorsh} As discussed above, this increase can be
attributed to the optical effect of the $\Delta_{\pi}$ gap, while
no clear effect of the $\Delta_{\sigma}$ gap is observable. In the
Al doped samples (panels b, c), \textit{RR} does not drastically
change with respect to the pure MgB$_{2}$ case. However, the small
edge shift, observed on going from x=0.0 to x=0.3, is not
consistent with the strong decrease of $\Delta_{\pi}$ (see Table
I). This suggests that the effect of the $\Delta_{\sigma}$ gap,
although not directly detectable, is not negligible in the
Al-doped samples. In the case of the irradiated sample (panel d),
on the contrary, remarkable changes in both shape
and intensity are observed. In particular, \textit{RR} starts
to increase around 90 cm$^{ - 1}$. We remark that this behavior
 cannot be explained by an optical effect of the
$\Delta_{\pi}$ gap, but must  be ascribed to an optical signature
of the $\Delta _{\sigma }$ gap.

To determine $\Gamma $  and $\Delta $  from the measured
\textit{RR} reflectivity ratios, it is necessary to model the
complex optical conductivity $\sigma$ in both  normal ($\sigma_N$)
and superconducting ($\sigma_S$) states. In the normal state,
$\sigma_N(\Omega,\Gamma;\omega)$ is provided by the Drude model,
which is well reliable below 100 cm$^{ - 1}$ since in this low
energy range neither optical phonon absorption or frequency
dependence of $\Gamma $ occur. In the superconducting state the
model developed by Zimmermann et al. \cite{zimm} was employed, as
in previous analysis of MgB$_2$ FIR spectra. \cite{pimen,jung}
 This model generalizes the MB theory and provides
$\sigma_S(\Omega^2/\Gamma,\Gamma,\Delta,T;\omega)$ at any given temperature $T<T_c$ for an arbitrary scattering
rate $\Gamma$. For a
conventional superconductor, test calculations showed that the
maximum in the resulting $R_{S}/R_{N}$ is evident when the ratio
$\Gamma / (2\Delta / \hbar)$ is at least of the order of 2. In the
MgB$_2$ case it is necessary to consider both $\pi$- and
$\sigma$-band contributions to the \textit{ab}-plane
(\textit{$\sigma $}$^{(ab)}$) and
\textit{c}-axis (\textit{$\sigma $}$^{(c)}$) conductivities.
For example, the normal state conductivities are given by:
\begin{equation}
\sigma_N^{(x)}(\omega) = \frac{[\Omega_{\sigma}^{(x)}]^2 /
\Gamma_{\sigma}}{1+ i\omega/ \Gamma_{\sigma}} +
\frac{[\Omega_{\pi}^{(x)}]^2 / \Gamma_{\pi}}{1+ i\omega/
\Gamma_{\pi}}
\end{equation}
\noindent with \textit{(x)=(a,b)} or \textit{(c)}. Since each
$\sigma ^{(x)}$ gives the reflectivity $R^{(x)}$, the reflectivity
$R$ of the polycrystalline sample can finally be evaluated as
$R$=(1/3)$ R^{(c)}$+ (2/3)$ R^{(a,b)}$, \cite{fudam} since it is
reasonable to assume that a large number of  randomly oriented
crystallites are impinged by the wide  IR beam.

The model $R_{S}/R_{N}$ curve was fitted to the experimental
\textit{RR} by employing only $\Delta_{\pi}, \Delta_{\sigma},
\Gamma_{\pi}$, and $\Gamma_{\sigma}$  as free parameters, since
 the plasma frequencies $\Omega$ cannot be
obtained by the present reflectivity measurements carried out over
a very limited spectral range. It is well known that in MgB$_{2}$
a strong discrepancy exists between the $\Omega $ calculated
 \cite{liu} and those observed in reflectance
measurements, \cite{fudam, kaind, peruc, jung} being the latter
typically smaller by a factor five than the former. A
discussion of this discrepancy, which is not yet fully
understood,\cite{fudam} is beyond the aim of the present paper.
The computed $\Omega $ values (see Table I) divided by a factor
5 were thus employed. For the irradiated sample D, the pure
MgB$_{2}$ $\Omega $ values were used, by assuming that irradiation
does not appreciably affect the electronic band structure, as can
be evinced by the weak reduction of $T_{C}$ in sample D (Table I).

Figure 1 shows that a satisfactory agreement is obtained between
the experimental data and the best-fit  curves for pure and
Al-doped samples (panels a, b, and c). The best-fit values of $\Gamma $ and
$\Delta $ are reported in Table I. The quoted errors on the
parameter values represent the range over which the parameters can
be varied without appreciably affecting the quality of the fit.
Note that in Al-doped samples the \textit{RR} curves were fitted
up to a maximum frequency corresponding to three times the
$\Delta_{\sigma }$ values provided by point-contact measurements
(see Table I). Approximations involved in the employed procedure
indeed are not expected to provide a reliable estimate of
$R_{S}/R_{N}$ at higher frequencies. \cite{vander} Also note that
in Table I and in the following of the paper the $\Gamma $ and
$\Delta $ values are given in meV units, for easier comparison
with previously reported values.  For pure MgB$_{2}$, the
$\Delta_{\pi }$ and $\Gamma_{\pi }$ values are determined with
good accuracy since the fit is very sensitive to the conduction
regime of the dirty $\pi $-band, which basically determines the
edge position of the $R_{S}/R_{N}$ increase. The $\Delta_{\sigma
}$ and $\Gamma_{\sigma }$ values for this sample are not reported
since these quantities cannot be unambiguously determined   for a
band in the clean limit. \cite{kam} Indeed, varying the values of
($\Delta_{\sigma }$, $\Gamma_{\sigma }$) within the range (5 meV, 2
meV)$\div$(12 meV, 10 meV) has no effect on the fit results for
$\Delta_{\pi }$ and $\Gamma_{\pi}$.
In the case of the Mg$_{1 - x}$Al$_{x}$B$_{2}$ samples,
$\Delta_{\pi}$ and $\Gamma_{\pi }$ are obtained with good
accuracy, as in pure MgB$_{2}$. Although with large uncertainties,
also $\Delta_{\sigma }$ and $\Gamma_{\sigma }$ are determined,
which indicates that the effect of the $\Delta_{\sigma }$ gap is
not negligible in the Al doped samples. It is remarkable that the
$\Delta _{\pi }$ values determined in this work are very close to
those directly measured with point-contact spectroscopy (see Table
I), \cite{put1} and to those determined through specific-heat
measurements. \cite{put3} In addition the present work provides a
valuable determination of the Al-doping dependence of $\Gamma_{\pi
}$, which increases by a factor 3 when the Al-doping is increased
up to 30{\%}. 

For the neutron irradiated sample, Fig. 1d shows that a good
agreement is obtained between experimental \textit{RR} and
computed $R_{S}/R_{N}$. A small uncertainty affects all the
best-fit parameter values (see Table I) since in this case both
the $\pi$ and $\sigma$ bands  play an important role in
determining shape and absolute values of $R_{S}/R_{N}$. It is
worth to notice that the $\Delta_{\pi}$ and $\Delta_{\sigma}$
values (not directly measured in the irradiated sample) are very
close to those of pure MgB$_{2}$. This result was expected since
the $T_{C}$ of the irradiated sample is close to that of pure
MgB$_{2}$. On the other hand, the relaxation rates $\Gamma$ are
strongly increased with respect to the pure sample and,
noticeably, they are nearly the same in $\pi $ and $\sigma $
bands, showing that the disorder introduced by irradiation is not
preferentially located in Mg or B planes.

Since the present analysis does  provide a
self-consistent estimate of the effect of Al-doping or irradiation
on the $\Gamma $ and $\Delta $ values, we finally report in
Table I the resulting $\Gamma_{\pi}/2\Delta_{\pi }$ and
$\Gamma_{\sigma} / 2\Delta_{\sigma }$ ratios, which characterize the
conduction regime of the $\pi $- and $\sigma $-bands. As for the
$\pi $-band, the $\Gamma_{\pi}/2\Delta_{\pi}$ values confirm the
dirty character of this band, which is increased by Al-doping (samples B and
C). It is worth to notice that the $\Gamma_{\pi}/2\Delta_{\pi}$ value in sample D is much higher than in pure
MgB$_{2}$, owing mainly due to the high $\Gamma_{\pi}$ value induced
by irradiation. As for the $\sigma$-band, no definitive assessment
can be made on its conduction regime for Al doped samples, owing
to the large errors affecting $\Gamma_{\sigma }/2\Delta_{\sigma}$
(see Table I). However, the present results suggest that, with
increasing Al doping, the $\sigma $-band also moves towards a
dirty regime, due to the combined effect of the large decrease of
$\Delta_{\sigma }$ and the small increase of $\Gamma_{\sigma}$.
This is consistent with a previous study of the Al-doping
dependence of the upper critical field B$_{C2}$ indicating the
dirty character of the $\sigma$-band in the x=0.3 sample.
\cite{put1} In the irradiated sample $\Gamma_{\sigma
}/2\Delta_{\sigma }>2$, owing mainly to the strong increase of
$\Gamma_{\sigma}$. This, combined with the high value of
$\Delta_{\sigma }$, makes well apparent the effect of the
$\sigma $-gap. We remark that this result allows to
clearly assess the dirty character of the $\sigma$-band in the
irradiated sample, in agreement with the upper critical field
analysis. \cite{put2}

In considering the results obtained for the $\sigma $ band in
Al-doped samples, it is worth to notice that the effect of the
$\sigma$-band on $R_{S}/R_{N}$ could be much more clearly
tested by probing the optical response of only the
\textit{ab}-plane, thus avoiding the \textit{c}-axis contribution which 
is dominated by carriers in the
dirty $\pi$-band. FIR measurements on the \textit{ab} face of
aluminum- and carbon-substituted single crystals \cite{karp} are
thus in program.

In conclusion, the dirty character of the $\pi$-band in all the
investigated MgB$_{2}$ samples is clearly evidenced by present results. In
Al-doped samples, no definitive assessment can be made on the conduction
regime of the $\sigma$-band, although obtained results suggest that it moves towards
a dirty regime with increasing Al doping. On the contrary, results obtained
for the irradiated sample show that the irradiation-induced disorder in both
the Mg and B planes drives the $\sigma$-band in the dirty regime without
significantly reducing $T_{C}$ and $\Delta_{\sigma}$, thus making the MgB$_{2}$
$\sigma$-gap observable for the first time in FIR measurements. Irradiation
therefore can provide an efficient method for obtaining a dirty MgB$_{2}$
superconductor with high $T_{C}$.

The authors thank P. Manfrinetti and A. Palenzona who prepared the samples
studied in this work.

\newcommand{\noopsort}[1]{} \newcommand{\printfirst}[2]{#1}
  \newcommand{\singleletter}[1]{#1} \newcommand{\switchargs}[2]{#2#1}

\begin{table}[p]
\begin {ruledtabular}
\caption{ Parameters characterizing the investigated MgB$_{2}$
(A), Mg$_{0.85}$Al$_{0.15}$B$_{2}$ (B),
Mg$_{0.7}$Al$_{0.3}$B$_{2}$ (C), and neutron irradiated MgB$_{2}$
(D) samples. Uncertainties on the last digit are reported in
brackets. $\ddag$ indicates  parameter values from Refs.\ 21,22,
 $\dag$
calculated plasma frequencies, and FIR parameter values obtained
from the fitting procedure (see text).} \vspace{0.5cm}
\begin{center}
\begin{tabular}{clcccc}

   &                           & A        &    B    &    C    &   D     \\
\hline\hline
$\ddag$&$T_C$(K)             & 39.0(2)  & 31(2)   & 24(3)   & 36.3(2) \\
&$\rho_{40K}$($\mu\Omega$cm)& 2.2(2)& 12(1) & 24(2)   & 20(2)   \\
&$\Delta_{\sigma}$(meV)& 7.4(5)   & 4(1)    & 2.0(2)  & -- \\
&$\Delta_{\pi}$(meV)    & 2.8(1)   & 1.7(2)  & 0.5(2)  & --   \\
\hline\hline
$\dag$&$\Omega_{\sigma}^{(a,b)}$(eV)&3.8 & 3.2     & 2.6     & --   \\
&$\Omega_{\sigma}^{(c)}$ & 0.7      & 0.7     & 0.7     & -- \\
&$\Omega_{\pi}^{(a,b)}$ & 5.5      & 5.9     &    6.0  & --   \\
&$\Omega_{\pi}^{(c)}$   & 6.4      &  6.8    &   6.9   & --\\
\hline\hline

  FIR   &$\Delta_{\sigma}$(meV)       & --     & 3.0(8)  & 3(1)    & 6.6(6)\\
      &$\Gamma_{\sigma}$       & --     & 9(3)    & 11(4)   & 31(4) \\
      &$\Delta_{\pi}$          & 2.4(2)   & 1.7(4)  & 1.2(3)  & 2.2(2)\\
      &$\Gamma_{\pi}$          & 11(1)    & 19(2)   & 27(3)   & 31(3) \\
\hline
 &$\Gamma_{\sigma}/2\Delta_{\sigma}$ &\text{--}&\text{1.6(5)}&\text{2.0(7)}& \text{2.4(3)} \\
 &$\Gamma_{\pi}/2\Delta_{\pi}$ &\text{2.4(3)}& \text{5.4(8)}  & \text{11(2)}   &\text{7.0(8)}\\
\end{tabular}
\end{center}
\end {ruledtabular}
\label{tablesummary}
\end{table}

\begin{figure}[p]
\input{epsf}
\epsfxsize 9cm
\centerline{\epsfbox{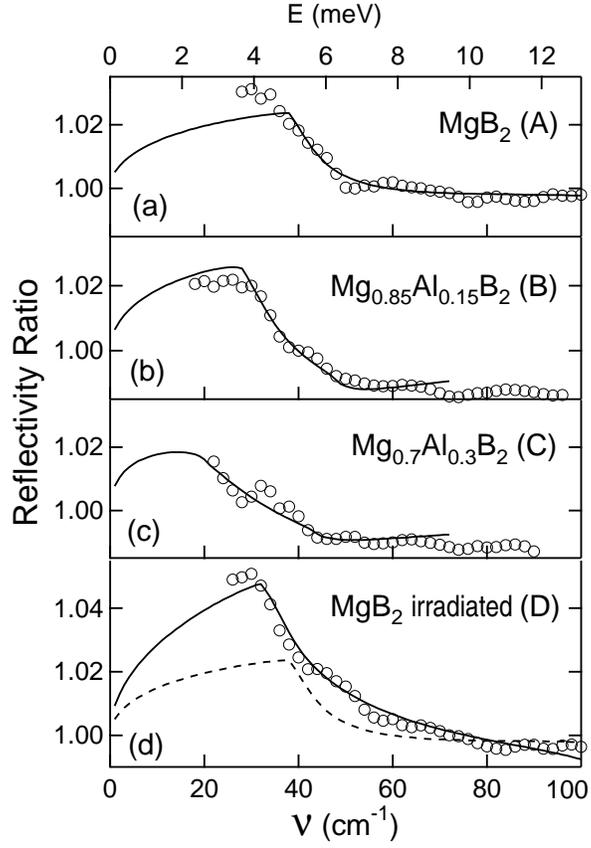}} \caption[~]{Measured \textit{RR}
(open symbols) and best-fit $R_{S}/R_{N}$ (full lines)
reflectivity ratios. In panel (d) the best-fit $R_{S}/R_{N}$ of
pure MgB$_2$ (panel a) is reported for comparison (dashed line).}
\label{fig1}
\end{figure}

\end{document}